\documentclass{article}

\usepackage[psamsfonts]{amssymb}
\usepackage{amsmath}
\usepackage{cite}
\usepackage{bm}
\usepackage{graphicx}
\usepackage{yhmath}
\usepackage{bbm}

\def\d{\mathrm{d}}

\def\id{\mathbf{1}}

\def\ir{\mathrm{i}}

\begin{document}
\begin{titlepage}
\noindent{\large\textbf{Loop diagrams in space with SU(2) fuzziness}}

\vskip 1 cm

\begin{center}
{Haniyeh Komaie-Moghaddam{\footnote {haniyeh.moghadam@gmail.com}}\\
Mohammad~Khorrami{\footnote {mamwad@mailaps.org}}\\
Amir~H.~Fatollahi{\footnote {ahfatol@gmail.com}}  } \vskip 10 mm
\textit{ Department of Physics, Alzahra University, Tehran
1993891167, Iran. }

\end{center}

\vspace{0.5cm}

\begin{abstract}
\noindent The structure of loop corrections is examined in a
scalar field theory on a three dimensional space whose spatial
coordinates are noncommutative and satisfy SU(2) Lie algebra. In
particular, the 2- and 4-point functions in $\phi^4$ scalar theory
are calculated at the 1-loop order. The theory is UV-finite as the
momentum space is compact. It is shown that the non-planar
corrections are proportional to a one dimensional
$\delta$-function, rather than a three dimensional one, so that in
transition rates only the planar corrections contribute.
\end{abstract}
\end{titlepage}

\section{Introduction}
In recent years there has been considerable interest in quantum
field theories on noncommutative spaces. This was to a large
extent motivated by the observation that this kind of field
theories arise in the zero-slope limit of the open string theory
in the presence of a constant B-field background
\cite{9908142,99-2,99-3,99-4}. In this case the coordinates
satisfy the canonical relation
\begin{equation}\label{kkf.1}
[\widehat x_\mu,\widehat x_\nu]=\ir\,\theta_{\mu \nu}\,\id,
\end{equation}
in which $\theta$ is an antisymmetric constant tensor and $\id$
represents the unit operator. The theoretical and phenomenological
implications of such noncommutative coordinates have been
extensively studied; see \cite{reviewnc}.

One direction to extend studies on noncommutative spaces is to
consider spaces where the commutators of the coordinates are not
constants. Examples of this kind are the noncommutative cylinder
and the $q$-deformed plane \cite{chai}, the so-called
$\kappa$-Poincar\'{e} algebra \cite{majid,ruegg,amelino,kappa},
and linear noncommutativity of the Lie algebra type \cite{wess}.
In the latter it is supposed that the dimensionless spatial
positions operators satisfy the commutation relations of a Lie
algebra \cite{wess}:
\begin{equation}\label{kkf.2}
[\widehat x_a,\widehat x_b]= f^c{}_{a\, b}\,\widehat x_c,
\end{equation}
where $f^c{}_{a\,b}$'s are structure constants of a Lie algebra.
One example of this kind is the algebra SO(3), or SU(2).
A special case of this is the so called fuzzy sphere \cite{madore,presnaj},
where an irreducible representation of the position operators is
used which makes the Casimir of the algebra,
$(\widehat x_1)^2+(\widehat x_2)^2+(\widehat x_3)^2$,
a multiple of the identity operator (a constant, hence the name
sphere). One can consider the square root of this Casimir as the
radius of the fuzzy sphere. This is, however, a noncommutative
version of a two-dimensional space (sphere).

In \cite{0612013,fakE1} a model was introduced in which the
representation was not restricted to an irreducible one, instead
the whole group was employed. In particular the regular
representation of the group, which contains all representations,
was considered. As a consequence in such models one is dealing
with the whole space, rather than a sub-space, like the case of
fuzzy sphere as a 2-dimensional surface. In \cite{0612013} the basic ingredients
for calculus on a linear fuzzy space, as well as the basic notions
for a field theory on such a space, were introduced. In
\cite{fakE1} the basic elements for calculating the matrix
elements corresponding to transition between initial and final
states were discussed. There the contributions of lowest order
(tree level) perturbative expansion of amplitudes were presented
for a self-interacting scalar field theory. The models based on
the regular representations of SU(2) was treated in more detail,
giving the explicit form of the tools and notions introduced in
their general forms \cite{0612013,fakE1}.

As mentioned in \cite{0612013,fakE1}, one of the features of
models based on linear fuzziness of Lie algebra type is that these
theories are free from any ultraviolet divergences if the
corresponding Lie group is compact. In fact one can consider the
momenta as the coordinates of the group, so that the space of the
corresponding momenta is compact iff the group is compact. One
important implication of the elimination of the ultraviolet
divergences would be that there will be no place for the so called
UV/IR mixing effect \cite{MRS}, which is known to be a common
feature of the models based on canonical noncommutativity, the
algebra (\ref{kkf.1}).

The purpose of the present work is to examine the structure of
the field theory amplitudes at loop order. Here we consider
a scalar field theory with $\phi^4$ interaction. In particular
we consider one-loop corrections to 2- and 4-point functions in
this theory. The field theory on a 2+1 spacetime whose coordinates
satisfy the Lie algebra of SO(2,1) was studied in \cite{naoki}.
Due to non-compactness of the group in this case, the UV-divergences
are present at loop level \cite{naoki}.

The scheme of the rest of this paper is the following. In section
2, a brief review is given on basic elements of a field theory on
a noncommutative space of SU(2) algebra type. In sections 3 and 4
the calculation of 2- and 4-point functions are presented,
respectively. Section 5 is devoted to concluding remarks; in
particular, it is discussed how only the planar sector of the loop
corrections contribute to the amplitudes.

\section{Field theory on space with SU(2) fuzziness}
In \cite{0612013,fakE1} a model was investigated in a 3+1
dimensional space-time the dimensionless spatial position
operators of which are generators of a \textit{regular}
representation of the SU(2) algebra, that is
\begin{equation}\label{kkf.3}
[\widehat x_a,\widehat x_b]=\epsilon^c{}_{a\,b}\,\widehat x_c.
\end{equation}
As it was discussed in \cite{0612013}, one can use the group
algebra as the analogue of functions defined on ordinary space,
with group elements $U=\exp(\ell\,k^a\,\widehat x_a)$ as the
analogues of $\exp(\ir\,\mathbf{k}\cdot\mathbf{x})$, which are a
basis for the functions defined on the space. In both cases
$\mathbf{k}$ is an ordinary vector with
$\mathbf{k}=(k^1,k^2,k^3)$. That is the components of $\mathbf{k}$
are commuting numbers. In the case of noncommutative space, $\ell$
is a length parameter, and the vector $\mathbf{k}$ is restricted
to a ball of radius $(2\,\pi/\ell)$, with all points of the
boundary identified to a single point. The manifold of
$\mathbf{k}$ is in fact a 3-sphere. $\mathbf{k}$ can be thought of
as the momentum of a particle. The left-right-invariant Haar
measure is
\begin{equation}\label{kkf.4}
\d U=\frac{\sin^2(\ell\,k/2)}{(\ell\, k/2)^2}\,\frac{\d^3 k}{(2\,\pi)^3},
\end{equation}
where $k:=|\mathbf{k}|$. The integration region for the
coordinates is $\displaystyle{k\leq 2\,\pi/\ell}$. We mention that
near the origin ($k\ll \ell^{-1}$) the measure is simply $\d^3
k/(2\,\pi)^3$, as it should be. The action of a scalar model with
quartic interaction in Fourier space of spatial directions is
given by
\begin{align}\label{kkf.5}
S=\int\d t\Bigg\{\frac{1}{2}&\int\d U_1\,\d
U_2\;\left[\dot\phi(U_1)\,\dot\phi(U_2)+
\phi(U_1)\,O(U_2)\,\phi(U_2)\right]\,\delta(U_1\,U_2)\cr
-&\frac{g}{4!}\int \Big[\prod_{j=1}^4 \d U_j \Big]
\;\phi(U_1)\,\phi(U_2)\,\phi(U_3)\,\phi(U_4)
\,\delta(U_1\,U_2\,U_3\,U_4)\Bigg\},
\end{align}
in which $\dot{\phi}$ is the time derivative of $\phi$. In the
above,
\begin{equation}\label{kkf.6}
O(U)=c\,\chi_\lambda(U+U^{-1}-2\,\id)-m^2,
\end{equation}
where $c$ and $m$ are constants, and $\chi_\lambda$ is the
character in the representation $\lambda$. It is shown that by a
proper choice of constant $c$, near the origin $O(U)\approx - k^2
-m^2$, as it is the case in the ordinary space. The
$\delta$-distribution appearing above is simply defined through
\begin{equation}\label{kkf.7}
\int\d U\;\delta(U)\,f(U):=f(\id),
\end{equation}
where $\id$ is the identity element of the group. It is easy to
see that this delta distribution is invariant under similarity
transformations, as well as inversion of the argument:
\begin{align}\label{kkf.8}
\delta(V\,U\,V^{-1})=&\,\delta(U),\nonumber\\
\delta(U^{-1})=&\,\delta(U).
\end{align}
The first relation shows that if the argument of the delta is a
product of group elements, then any cyclic permutation of these
elements leaves the delta unchanged. It is also seen that near the
origin ($k\ll \ell^{-1}$),
\begin{equation}\label{kkf.9}
\delta(U_1\,\cdots\,U_l)\approx (2\,\pi)^3\, \delta^3
(\mathbf{k}_1+\cdots+\mathbf{k}_l),
\end{equation}
which ensures an approximate momentum conservation. The exact
conservation law, however, is that at each vertex the product of
incoming group elements should be unity. For the case of a
3-leg vertex, one can write this condition as
\begin{equation}\label{kkf.10}
\exp(\ell\,k_1^a\,\widehat x_a)\,\exp(\ell\,k_2^a\,\widehat x_a)\,
\exp(\ell\,k_3^a\,\widehat x_a)=\id.
\end{equation}
It is convenient to define
\begin{equation}\label{kkf.11}
\exp(\ell\,k_1^a\,\widehat x_a)\,\exp(\ell\,k_2^a\,\widehat x_a)=:
\exp[\ell\,\gamma^a(\mathbf{k}_1,\mathbf{k}_2)\,\widehat x_a],
\end{equation}
where the function $\boldsymbol{\gamma}$ can be shown to enjoy the
properties
\begin{align}\label{kkf.12}
\boldsymbol{\gamma}[\mathbf{k}_1,
\boldsymbol{\gamma}(\mathbf{k}_2,\mathbf{k}_3)]=&
\boldsymbol{\gamma}[\boldsymbol{\gamma}(\mathbf{k}_1,\mathbf{k}_2),
\mathbf{k}_3],\\ \label{kkf.13}
\boldsymbol{\gamma}(-\mathbf{k}_1,-\mathbf{k}_2)=&
-\boldsymbol{\gamma}(\mathbf{k}_2,\mathbf{k}_1),\\ \label{kkf.14}
\boldsymbol{\gamma}(\mathbf{k},-\mathbf{k})=&0.
\end{align}
So that (\ref{kkf.10}) can be expressed by one of the three
equivalent forms
\begin{align}\label{kkf.15}
\mathbf{k}_3=&-\boldsymbol{\gamma}(\mathbf{k}_1,\mathbf{k}_2),\cr
\mathbf{k}_2=&-\boldsymbol{\gamma}(\mathbf{k}_3,\mathbf{k}_1),\cr
\mathbf{k}_1=&-\boldsymbol{\gamma}(\mathbf{k}_2,\mathbf{k}_3).
\end{align}
The explicit form of
$\boldsymbol{\gamma}(\mathbf{k}_1,\mathbf{k}_2)$ is obtained from
\begin{align}\label{kkf.16}
\cos\frac{\ell\,\gamma}{2}=&
~\cos\frac{\ell\,k_1}{2}\,\cos\frac{\ell\,k_2}{2}-
\mathbf{\hat{k}}_1\cdot\mathbf{\hat{k}}_2\,
\sin\frac{\ell\,k_1}{2}\,\sin\frac{\ell\,k_2}{2},\cr
\mathbf{\hat{\boldsymbol{\gamma}}}\,\sin\frac{\ell\,\gamma}{2}=&
~\mathbf{\hat{k}}_1\times\mathbf{\hat{k}}_2\,
\sin\frac{\ell\,k_1}{2}\,\sin\frac{\ell\,k_2}{2}\cr &+
\mathbf{\hat{k}}_1\,\sin\frac{\ell\,k_1}{2}\,\cos\frac{\ell\,k_2}{2}+
\mathbf{\hat{k}}_2\,\sin\frac{\ell\,k_2}{2}\,\cos\frac{\ell\,k_1}{2}.
\end{align}
It is easy to see that in the limit $\ell\to 0$,
$\boldsymbol{\gamma}$ tends to $\mathbf{k}_1+\mathbf{k}_2$, as
expected.

For field theoretical purposes it is convenient to have the action
(\ref{kkf.5}) in the whole (space and time) Fourier space:
\begin{align}\label{kkf.17}
S=&\frac{1}{2}\int\frac{\d\omega_1\,\d
U_1}{2\,\pi}\,\frac{\d\omega_2\,\d U_2}{2\,\pi}
~[2\,\pi\,\delta(\omega_1+\omega_2)\,\delta(U_1\,U_2)]\cr &~\times
\left[-\omega_1\,\omega_2\,
\check\phi(U_1,\omega_1)\,\check\phi(U_2,\omega_2)+
\check\phi(U_1,\omega_1)\,O(U_2)\,\check\phi(U_2,\omega_2)\right]\cr
-&\frac{g}{4!}\int\left[\prod_{j=1}^4 \frac{\d\omega_j\,\d
U_j}{2\,\pi}\;\check\phi(U_j,\omega_j)\right]
\,[2\,\pi\,\delta(\omega_1+\cdots+\omega_4)\,\delta(U_1\cdots
U_4)],
\end{align}
in which $\check\phi(U,\omega)$ is the Fourier component. The
first two terms represent a free action, with the propagator
\begin{equation}\label{kkf.18}
\check\Delta(\omega,U):=\frac{\ir\,\hbar}{\omega^2+O(U)}.
\end{equation}
Putting the denominator of this propagator equal to zero gives the
relation between $\omega$ and $U$ for free particles (the
mass-shell condition). The third term contains interactions. Any
Feynman graph would consist of propagators, and $4$-line vertices
to which one assigns the fundamental vertex
\begin{align}\label{kkf.19}
\mathcal{V}_{[1234]}:=\frac{g}{\ir\,\hbar\,4!}\,2\,\pi\,
\delta(\omega_1+\cdots+\omega_4)\,\sum_{\Pi}\delta(U_{\Pi(1)}
\cdots U_{\Pi(4)}),
\end{align}
where the summation runs over all permutations. In practice, due
to cyclic symmetry of $\delta$'s arguments mentioned earlier,
permutations which are different up to a cyclic change just come
in sum with a proper weight, so we have
\begin{align}\label{kkf.20}
\!\mathcal{V}_{[1234]}=&\frac{g}{\ir\,\hbar\,6}\,2\,\pi\,
\delta(\omega_1+\cdots+\omega_4)\, \Big[
\delta(U_1U_2U_3U_4)+\delta(U_1U_2U_4U_3)\cr
&+\delta(U_1U_3U_2U_4)+\delta(U_1U_3U_4U_2)+
\delta(U_1U_4U_2U_3)+\delta(U_1U_4U_3U_2)\Big].\qquad
\end{align}
Also, for any internal line there is an integration over $U$ and
$\omega$, with the measure $\d\omega\,\d U/(2\,\pi)$. As the group
is assumed to be compact, the integration over the group is
integration over a compact volume.

It is worth to mention a crucial difference between the way that
$\delta$-functions appear in our model and in models defined on
ordinary spaces. Here, as mentioned above, each possible ordering
of legs of a vertex comes with a different $\delta$, except the
cases that two orderings are different up to a cyclic permutation.
This is in contrast to models on ordinary space, in which all
possible orderings have the common factor of one single
$\delta\big( \sum \mathbf{k}_i\big)$, representing the momentum
conservation in that vertex. Similar to above observation about
the appearance of $\delta$-functions has been made in theories
defined on $\kappa$-deformed spaces, pointed in the Introduction.
In these theories, the ordinary summation of momenta in each
vertex is replaced with a new rule of summation, occasionally
called as doted-sum ($\dot{+}$) \cite{amelino}. This new sum, in
contrast to the ordinary sum, is non-Abelian, and as a
consequence, the $\delta$'s coming with each possible ordering of
legs are different \cite{amelino,kappa}.

Once given by the Feynman rules one can calculate the loop
corrections. In following we choose $\lambda=\frac{1}{2}$ in
(\ref{kkf.6}), so that the propagator has the explicit form
\begin{equation}\label{kkf.21}
\check\Delta(\omega,\mathbf{k})=
\frac{\ir\,\hbar} {\omega^2-\displaystyle{\frac{16}{\ell^2}\,
\sin^2\frac{\ell\, k}{4}}-m^2}.
\end{equation}

\section{1-loop correction of the 2-point function}
The 2-point function has two external legs, one incoming
$(\omega_1,\mathbf{k}_1)$, the other outgoing
$(-\omega_2,-\mathbf{k}_2)$. The 1-loop correction is simply the
fundamental vertex-function (\ref{kkf.20}), contracted on legs $3$
and $4$, with proper symmetry factors:
\begin{align}\label{kkf.22}
\Gamma^{(2)}_{\rm 1-loop}=& \frac{1}{2}\int\d U_3\,\d U_4\;
\delta(U_3\,U_4)\int
\frac{\d\omega_3}{2\pi}\,\frac{\d\omega_4}{2\pi}\;
2\,\pi\,\delta(\omega_3+\omega_4)\,\mathcal{V}_{[1234]},\nonumber\\
=& \frac{g}{\ir\,\hbar\,12}\,2\,\pi\,\delta(\omega_1-\omega_2)\,
\int \d U \int \frac{\d\omega}{2\pi}\;\frac{\ir\,\hbar}{\omega^2 +
O(U)}
\nonumber\\
& \times \left[ 4\, \delta(U_1\, U_2^{-1})+2\,\delta(U_1\,U\,
U_2^{-1}\,U^{-1})\right],
\end{align}
where the integrations on $\omega_4$ and $U_4$ have been
performed, and $\omega_3$ and $U_3$ have been denoted by $\omega$
and $U$, respectively. Also use has been made of the facts that
$\d U$ and $O(U)$ are invariant under the inversion of $U$. In the
above expression, one recognizes two parts: the so called planar
part, in which the delta contains no contribution from the loop
momentum $U$; and the so called nonplanar part, in which the delta
does contain loop momentum. Using the usual prescription
$\omega^2\to(\omega^2+\ir\,\varepsilon)$, the integration over
omega is performed. For the planar part, one obtains
\begin{align}\label{kkf.23}
\Gamma^{(2){\rm planar}}_{\rm 1-loop} =&-\frac{\ir
\,g}{6}\,2\,\pi\,\delta(\omega_1-\omega_2)\,\delta(U_1\,U_2^{-1})
\int \frac{\d U}{\sqrt{- O(U)}}\cr =&-\frac{\ir\, 8\,\pi\,
g}{3\,\ell^2}\,2\,\pi\,\delta(\omega_1-\omega_2)\,\delta(U_1\,
U_2^{-1})\int_0^{2\,\pi/\ell}\!\frac{\d k}
{(2\,\pi)^3}\frac{\sin^2(\ell\,k/2)}
{\sqrt{16\sin^2(\ell\,k/4)+m^2}}.\qquad\nonumber\\
\end{align}
The remaining integral can be expressed in terms of a
hypergeometric function of $_2F_1(a,b,c;z)$ type. It is seen that
the above expression is finite, as it was to be. The delta
distribution of group elements can also be written in the form
\begin{equation}\label{kkf.24}
\delta(U_1\,U_2^{-1}) = \frac{(2\,\pi)^3\,(\ell\,
k_2/2)^2}{\sin^2(\ell\,k_2/2)}
\,\delta^3(\mathbf{k}_1-\mathbf{k}_2).
\end{equation}
It is seen that planar contribution conserves momentum.

For the non-planar contribution, one has
\begin{align}\label{kkf.25}
\Gamma^{(2){\rm non\!-\!planar}}_{\rm 1-loop} =&-\frac{\ir
\,g}{12}\,2\,\pi\,\delta(\omega_1-\omega_2) \int \d
U\;\frac{\delta(U_1\,U\,U_2^{-1}\,U^{-1})}{\sqrt{- O(U)}}.
\end{align}
It is convenient to define $\mathbf{k}'_2$ through
\begin{align}\label{kkf.26}
U(\mathbf{k}'_2):=&\,U(\mathbf{k})\,U_2\,U^{-1}(\mathbf{k}),\nonumber\\
\mathbf{k}'_2=&\,\mathbf{k}_2\,\cos(\ell\,k)+
\hat{\mathbf{k}}\,(\hat{\mathbf{k}}\cdot\mathbf{k}_2)\,
[1-\cos(\ell\, k)] + \mathbf{k}_2\times\hat{\mathbf{k}}\,
\sin(\ell\, k).
\end{align}
In fact $\mathbf{k}'_2$ is nothing but $\mathbf{k}_2$ rotated by
the angle $\ell\,k$ around $\hat{\mathbf{k}}$. So,
\begin{equation}\label{kkf.27}
k'_2=k_2,
\end{equation}
and
\begin{align}\label{kkf.28}
\delta(U_1\,U\,U_2^{-1}\,U^{-1})=&\,\delta(U_1\,U'^{-1}_{2}),\nonumber\\
=&\,\frac{(2\,\pi)^3(\ell\, k_2/2)^2}{\sin^2(\ell\,k_2/2)}
\,\delta^3(\mathbf{k}_1-\mathbf{k}'_2).
\end{align}
It would be helpful to express this delta in terms of spherical
coordinates:
\begin{equation}\label{kkf.29}
\delta^3(\mathbf{k}_1-\mathbf{k}'_2)=\frac{1}{k_1^2}\,
\delta(k_1-k_2)
\,\delta(\cos\theta_1-\cos\theta'_2)\,\delta(\phi_1-\phi'_2).
\end{equation}
Without loss of generality, one can put the 3rd direction on
$\mathbf{k}_2$. So,
\begin{align}\label{kkf.30}
\mathbf{k}_2=&\,k_2\,\hat{\mathbf{z}},\nonumber\\
\cos\theta'_2=&\, \cos(\ell\,k) + \cos^2\theta\,[1-\cos (\ell\, k)]\nonumber\\
\phi'_2=&\,\phi-\tan^{-1}\left\{\frac{\sin (\ell\,
k)}{\cos\theta\,[\cos (\ell\, k)-1]}\right\},
\end{align}
where $(k,\theta,\phi)$ are the spherical coordinates of
$\mathbf{k}$. Since $O(U)$ is independent of $\theta$ and $\phi$,
one can use
\begin{align}\label{kkf.31}
I_1:=&\,\int_{-1}^1\d(\cos\theta)\;\delta\left\{\cos\theta_1 -
\cos(\ell\,k)-\cos^2\theta\,[1-\cos (\ell\,
k)]\right\},\nonumber\\
=&\,
\begin{cases}\left\{[1-\cos(\ell\,
k)]\,[\cos\theta_1-\cos(\ell\, k)]\right\}^{-1/2},& \theta_1 \leq
\ell\,k\leq 2\,\pi-\theta_1\\
0,& \textrm{otherwise}\end{cases}\nonumber\\
I_2:=&\int \d\phi \;\delta(\phi_1-\phi'_2)=1.
\end{align}
One then arrives at
\begin{align}\label{kkf.32}
\Gamma^{(2){\rm non\!-\!planar}}_{\rm 1-loop} =&-\frac{\ir
\,g}{12}\,2\,\pi\,\delta(\omega_1-\omega_2)\,\delta(k_1-k_2)\,
\frac{(2\,\pi)^3\,\ell^2/4}{\sin^2(\ell\,k_2/2)}\nonumber\\
&\times \int_{\theta_1/\ell}^{(2\,\pi-\theta_1)/\ell} \frac{\d
k}{(2\,\pi)^3}\,\frac{\sin^2(\ell,k/2)}
{\sqrt{16\sin^2(\ell\,k/4)+m^2}}\nonumber\\
&\quad\times \frac{1}{\sqrt{[1-\cos(\ell\,
k)]\,[\cos\theta_1-\cos(\ell\, k)]}}.
\end{align}
Putting $\cos^{-1}(\hat{\mathbf{k}}_1\cdot\hat{\mathbf{k}}_2)$
instead of $\theta_1$, an expression is obtained that has no
particular choice of coordinates. The above expression is finite,
hence no appearance of UV-divergences in either planar or
non-planar diagrams. However, in the above expression one has a
one-dimensional delta $\delta(k_1-k_2)$, instead of the three
dimensional delta $\delta^3(\mathbf{k}_1-\mathbf{k}_2)$ appearing
in the planar part. So the non-planar part does not leave the
momentum vector but only its length conserved. So it is possible
that the direction of the momentum of a self-interacting particle
changes. This is no surprise, as in this theory momentum
conservation is just an approximate law. Similar observations on
the change of momentum through non-planar loop corrections to
2-point functions have been reported in $\kappa$-Poincar\'{e}
theories \cite{amelino,kappa}

\section{1-loop correction of the 4-point function}
The 4-point function has four external legs: two incomings
$(\omega_1,\mathbf{k}_1)$ and $(\omega_2,\mathbf{k}_2)$, and two
outgoings $(-\omega_3,-\mathbf{k}_3)$ and
$(-\omega_4,-\mathbf{k}_4)$. The contributions come from three
channels, the so-called s-, t-, and u-channels. Here only the
s-channel is investigated, as the two others can be obtained
similarly. One has
\begin{align}\label{kkf.33}
\Gamma^{(4){\rm [s]}}_{\rm 1-loop} =&\,\frac{1}{2} \int \d U \d U'
\int \frac{\d\omega}{2\pi}\frac{\d\omega'}{2\pi}\nonumber\\
&\times \frac{\ir\,\hbar}{\omega^2 +
O(U)}~\frac{\ir\,\hbar}{\omega'^2 +
O(U')}\,\mathcal{V}_{[12U^{-1}U'^{-1}]}\,\mathcal{V}_{[UU',-3,-4]}.
\end{align}
Using (\ref{kkf.20}), and performing the $\omega'$-integration,
one would get
\begin{align}\label{kkf.34}
\Gamma^{(4){\rm [s]}}_{\rm 1-loop} =&\,\frac{1}{2}
\left(\frac{g}{6\,\ir\,\hbar}\right)^2
2\pi\,\delta(\omega_1+\omega_2-\omega_3-\omega_4) \int \d U \d U' \int
\frac{\d\omega}{2\,\pi}\nonumber\\
&\times \frac{\ir\,\hbar}{\omega^2 +
O(U)}\,\frac{\ir\,\hbar}{(\omega_{\rm s}-\omega)^2 +
O(U')}\nonumber\\
&\times[\delta(U_1\,U_2\,U^{-1}\,U'^{-1})+
\delta(U_1\,U_2\,U'^{-1}\,U^{-1})+
\delta(U_1\,U^{-1}\,U_2\,U'^{-1})\nonumber\\
&+\delta(U_1\,U'^{-1}\,U_2\,U^{-1})+
\delta(U_2\,U_1\,U^{-1}\,U'^{-1})+
\delta(U_2\,U_1\,U'^{-1}\,U^{-1})]\nonumber\\
&\times[\delta(U_3^{-1}\,U_4^{-1}\,U\,U')+
\delta(U_3^{-1}\,U_4^{-1}\,U'\,U)+
\delta(U_3^{-1}\,U\,U_4^{-1}\,U')\nonumber\\
&+\delta(U_3^{-1}\,U'\,U_4^{-1}\,U)+
\delta(U_4^{-1}\,U_3^{-1}\,U\,U')+
\delta(U_4^{-1}\,U_3^{-1}\,U'\,U)],
\end{align}
in which
\begin{equation}\label{kkf.35}
\omega_{\rm s}:=\omega_1+\omega_2.
\end{equation}
It would be convenient to define the followings
\begin{align}\label{kkf.36}
U'_a:=&\,U\,U_a\,U^{-1},\nonumber\\
U''_a:=&\,U^{-1}\,U_a\,U.
\end{align}
Performing the $U'$-integration, and the change
$\omega\to\omega-\omega_{\rm s}$, one arrives at
\begin{align}\label{kkf.37}
\Gamma^{(4){\rm [s]}}_{\rm 1-loop} =&
\frac{1}{2}\,\left(\frac{g}{6}\right)^2
2\pi\,\delta(\omega_1+\omega_2-\omega_3-\omega_4) \int \d U  \int
\frac{\d\omega}{2\,\pi}\;\frac{1}{(\omega-\omega_{\rm s})^2 +
O(U)}\nonumber\\
&\times\Bigg\{\frac{1}{\omega^2 + O(U_1\,U_2\,U^{-1})}\,[
2\,\delta(U_1\,U_2\,U_3^{-1}\,U_4^{-1})+
2\,\delta(U_1\,U_2\,U_4^{-1}\,U_3^{-1})\nonumber\\
&+\delta(U_1\,U_2\,U''^{-1}_{3}\,U_4^{-1})+
\delta(U_1\,U_2\,U''^{-1}_4\,U_3^{-1})
+\delta(U_1\,U_2\,U_3^{-1}\,U'^{-1}_4)\nonumber\\
&+\delta(U_1\,U_2\,U_4^{-1}\,U'^{-1}_3)+
\delta(U_1\,U_3^{-1}\,U_4^{-1}\,U'_2)
+\delta(U''_1\,U_3^{-1}\,U_4^{-1}\,U_2)\nonumber\\
&+\delta(U_1\,U_4^{-1}\,U_3^{-1}\,U'_2)
+\delta(U''_1\,U_4^{-1}\,U_3^{-1}\,U_2)\nonumber\\
&+\delta(U'_1\,U'_2\,U_4^{-1}\,U_3^{-1})
+\delta(U''_1\,U''_2\,U_3^{-1}\,U_4^{-1})
+\delta(U''_1\,U''_2\,U_4^{-1}\,U_3^{-1})\nonumber\\
&+\delta(U_1\,U_3^{-1}\,U'^{-1}_4\,U'_2)
+\delta(U_1\,U_4^{-1}\,U'^{-1}_3\,U'_2)
+\delta(U'_1\,U'_2\,U_3^{-1}\,U_4^{-1})]\nonumber\\
&+\frac{1}{\omega^2 + O(U_2\,U_1\,U^{-1})}\,
[2\,\delta(U_2\,U_1\,U_3^{-1}\,U_4^{-1})
+2\,\delta(U_2\,U_1\,U_4^{-1}\,U_3^{-1})\nonumber\\
&+\delta(U'_1\,U_2\,U_3^{-1}\,U_4^{-1})
+\delta(U_1\,U''_2\,U_3^{-1}\,U_4^{-1})
+\delta(U'_1\,U_2\,U_4^{-1}\,U_3^{-1})\nonumber\\
&+\delta(U_1\,U''_2\,U_4^{-1}\,U_3^{-1})
+\delta(U_2\,U_1\,U''^{-1}_3\,U_4^{-1})
+\delta(U_2\,U_1\,U''^{-1}_4\,U_3^{-1})\nonumber\\
&+\delta(U_2\,U_1\,U_3^{-1}\,U'^{-1}_4)
+\delta(U_2\,U_1\,U_4^{-1}\,U'^{-1}_3)\nonumber\\
&+\delta(U''_2\,U''_1\,U_4^{-1}\,U_3^{-1})
+\delta(U'^{-1}_4\,U'_1\,U_2\,U_3^{-1})
+\delta(U'^{-1}_3\,U_1\,U_2\,U_4^{-1})\nonumber\\
&+\delta(U'_2\,U'_1\,U_3^{-1}\,U_4^{-1})
+\delta(U'_2\,U'_1\,U_4^{-1}\,U_3^{-1})
+\delta(U''_2\,U''_1\,U_3^{-1}\,U_4^{-1})]\Bigg\},
\end{align}
where use has been made of
\begin{equation}
O(ABC)=O(CAB).
\end{equation}
In the above expression, the delta's come in two way: those
without the loop variable, which correspond to the planar part;
and those containing the loop variable (those which contain $'$ or
$''$), which correspond to the non-planar part. In the planar
part, the delta's are simply brought out of the integral. So,
\begin{align}\label{kkf.39}
\Gamma^{(4){\rm [s]planar}}_{\rm 1-loop}= &\,\frac{g^2}{72\pi}\,
\delta(\omega_{\rm s}-\omega_3-\omega_4)\,\Bigg\{
[\delta(U_1\,U_2\,U^{-1}_3\,U^{-1}_4)
+\delta(U_1\,U_2\,U^{-1}_4\,U^{-1}_3)]\nonumber\\
&\times\int \d U \int \frac{\d\omega}{[(\omega-\omega_{\rm s})^2 +
O(U)]\,[\omega^2 + O(U_1\,U_2\,U^{-1})]}\nonumber\\
&+[\delta(U_2\,U_1\,U^{-1}_3\,U^{-1}_4)
+\delta(U_2\,U_1\,U^{-1}_4\,U^{-1}_3)]\nonumber\\
&\times \int \d U \int \frac{\d\omega}{[(\omega-\omega_{\rm s})^2
+ O(U)]\,[\omega^2 + O(U_2\,U_1\,U^{-1})]}\Bigg\}.
\end{align}
Again the contribution of the planar part is proportional to
three-dimensional delta's.

One can proceed to bring the above expressions in more simple
forms, though in this case the integrand does not just depend on
the length of momentum. As an easy example, one can consider the
case where the reactions takes place in the so called center of
mass frame:
\begin{equation}\label{kkf.40}
U_2=U_1^{-1}.
\end{equation}
One then arrives at
\begin{align}\label{kkf.41}
\Gamma^{(4){\rm [s]planar}}_{\rm 1-loop} =&
\,\frac{g^2}{18\pi}\,\delta(\omega_{\rm s}-\omega_3-\omega_4)
\,\delta(U_3U_4)\nonumber\\
&\times \int \d U \int \frac{\d\omega}{[(\omega-\omega_{\rm s})^2
+ O(U)]\,[\omega^2 + O(U)]}.
\end{align}

For the non-planar contribution, as it was the case with the
two-point function, one cannot factor out a three-dimensional
delta, as the loop variable is inside the argument of the delta's.
It is again obvious that both the planar and non-planar
contributions are finite.

\section{Concluding remarks}
The structure of loop corrections of a self-interacting field
theory on a three dimensional space whose spatial coordinates are
noncommutative and satisfy SU(2) Lie algebra was examined. The
examples of 1-loop 2- and 4-point functions were treated in more
detail as examples of loop corrections in such models. As the
momentum space of such models are compact, the theory is free from
UV divergences.

In the case of the 2-point function, while the planar part leaves
the momentum conserved, the non-planar only leaves the length of
the momentum conserved. In the case of the 4-point function, the
momentum is conserved neither by the planar part nor by the
non-planar part. Similar observations have already been done in
the case of the $\kappa$-Poincar\'{e} case \cite{amelino}.

One notable feature of the model is about the ways that the
$\delta$-functions appear in planar and non-planar sectors of the
theory. The planar contribution comes with a three dimensional
$\delta$, representing certain combinations of the external
momentums of the diagrams. In the non-planar case,
less-dimensional delta's remain. As a consequence, the $n$-point
function can always come schematically as below
\begin{align}\label{kkf.42}
\!\!\!\Gamma^{(n)} = 2\,\pi\,\delta\left(\sum_{\mathrm{i}
\,\&\,\mathrm{f}} \omega\right)\,\left[ \sum_\lambda
\Gamma^{(n)\,{\rm planar}}_\lambda\delta^3(\mathbf{v}_\lambda) +
\sum_{\mu}\Gamma^{(n)\,{\rm non-planar}}_{\mu}
\delta^{\alpha_\mu}(\mathbf{v}_{\mu}) \right],
\end{align}
where $\lambda$ and $\mu$ denote different orderings of external
legs. The vectors $\mathbf{v}_\lambda$ and $\mathbf{v}_\mu$ are
certain combinations of external momenta. Finally, the numbers
$\alpha_\mu$ are less than 3, meaning that the delta arising from
the non-planar part is less than three-dimensional.

Observables (like cross sections and decay rates) are proportional
to the square of $\Gamma^{(n)}$, in which there are terms
proportional to
$[\delta^\alpha(\mathbf{v})]\,[\delta^{\alpha'}(\mathbf{v}')]$,
where $\alpha$ or $\alpha'$ is equal to 3 for contributions from
the planar sector, and less than 3 for contributions from the
non-planar sector. It is seen that in such products, there arise
terms proportional to $\delta^\beta(\mathbf{0})$, such that
$\beta=3$ iff $\mathbf{v}=\mathbf{v}'$ and $\alpha=\alpha'=3$, and
$\beta<3$ otherwise. So $\delta^3(\mathbf{0})$ arises only in the
product of similar terms in the planar sector. As it was discussed
in \cite{fakE1}, to obtain transition rates these terms should be
multiplied by other factors including powers of the volume of the
space, which tends to $\delta^3(\mathbf{0})$ in the infinite
volume limit. These factors cancel one $\delta^3(\mathbf{0})$ from
the square, so that terms containing one $\delta^3(\mathbf{0})$
give finite contributions. Other terms vanish, and as terms
containing one $\delta^3(\mathbf{0})$ arise only from the planar
parts, the non-planar corrections do not contribute in the
probability of any transition rate. The only exception is for the 2-point
function, where the planar part has no contribution to the
self-scattering (spontaneous momentum change) and the scattering
is totally due to the non-planar part.

\vspace{.5cm}

\noindent\textbf{Acknowledgement}: This work was partially
supported by the research council of the Alzahra University.

\end{document}